\documentclass[aps,reprint,superscriptaddress,floatfix] {revtex4-1}
\usepackage{graphicx}
\usepackage{color}
\usepackage{float}
\usepackage[normalem]{ulem}

\begin{document}
\title{Strain-driven attenuation of superconductivity in heteroepitaxial perovskite/YBCO/perovskite thin films}

\author{H. Zhang}
\affiliation{Department of Physics, University of Toronto, Toronto, M5S1A7, Canada}

\author{A. Nguyen}
\affiliation{Department of Physics \& Astronomy, California State University Long Beach, Long Beach, 90840, USA}

\author{T. Gredig}
\affiliation{Department of Physics \& Astronomy, California State University Long Beach, Long Beach, 90840, USA}

\author{J. Y.T. Wei}
\affiliation{Department of Physics, University of Toronto, Toronto, M5S1A7, Canada}
\affiliation{Canadian Institute for Advanced Research, Toronto, M5G1Z8, Canada}

\begin{abstract} 

To distinguish between the effects of strain and magnetism on the superconductivity in $c$-axis La$_{2/3}$Ca$_{1/3}$MnO$_3$/YBa$_2$Cu$_3$O$_{7-\delta}$ (LCMO/YBCO) heterostructures, we study perovskite/YBCO/perovskite thin films using either ferromagnetic LCMO or paramagnetic LaNiO$_3$ (LNO) as perovskite.  For a lattice-symmetry matched comparison, we also use orthorhombic PrBa$_2$Cu$_3$O$_{7-\delta}$ (PBCO) in place of the pseudocubic perovskites. Unlike PBCO/YBCO/PBCO, both LCMO/YBCO/LCMO and LNO/YBCO/LNO trilayers show strong attenuation of the superconducting $T_c$ as YBCO layer thickness is reduced from 21.4 to 5.4 nm.  Our results indicate that heteroepitaxial strain, rather than long-range proximity effect, is responsible for the long length scales of $T_c$ attenuation observed in $c$-axis LCMO/YBCO heterostructures.

\end{abstract}

\maketitle

In thin-film heterostructures of complex oxides, heteroepitaxial strain due to interfacial lattice mismatch can significantly affect their electronic properties, by virtue of the sensitivity to bond lengths and angles \cite{markert, may}. Examples include enhancement of the superconducting critical temperature ($T_c$) in cuprates \cite{locquet,bozovic}, variation of the Mott gap size in iridates \cite{serrao}, modulation of the conductivity in nickelates \cite{burriel}, and tunable carrier mobility at aluminate/titanite interfaces \cite{jalan}. Although the heteroepitaxial strain in oxide thin films can be relieved by dislocations, cracking and formation of vacancies \cite{matthews, thouless, klie, aschauer}, studies of various oxides including cuprates and manganites have shown that the strain can extend well into the film, as far as $\sim$ 200 nm from the interface \cite{meyer, peng, peng-apl, wang}. In addition to lattice-parameter mismatch, lattice-symmetry mismatch is also known to affect the physical properties of oxide heterostructures via anisotropic strain \cite{tan}.

In $c$-axis thin-film heterostructures comprising the high-$T_c$ cuprate $\mathrm{YBa_2Cu_3O_{7-\delta}}$ (YBCO) and the half-metallic manganite $\mathrm{La_{2/3}Ca_{1/3}MnO_3}$ (LCMO), heteroepitaxial strain has been shown to induce CuO intergrowths in the YBCO layer \cite{zhang}. Attributed to lattice-symmetry mismatch between the orthorhombic YBCO and the pseudocubic LCMO (see Table 1), these CuO intergrowths form nanoscale phase inhomogeneity which can attenuate $T_c$.  This strain-based mechanism of $T_c$ attenuation has crucial implications for the study of ferromagnet/superconductor (F/S) proximity effect in these heterostructures \cite{goldman, sefrioui, soltan, pang, pena, chakhalian, freeland, dybko, hoppler, fridman, visani12, kalcheim2, visani15, gray}, as the purportedly long range ($\sim$ 10 - 20 nm) of this proximity effect was inferred from the length scale of $T_c$ attenuation versus either decreasing YBCO or increasing LCMO layer thickness \cite{sefrioui, soltan, pang, pena}.  Since the observed length scales are within the extent of heteroepitaxial strain, it is questionable whether they can simply be identified with the proximity effect.  In fact, although a long-range F/S proximity effect is expected from theory based on spin-triplet pairing \cite{buzdin, bergeret, asano, fominov, linder, hoegl}, the actual range of the proximity effect in $c$-axis LCMO/YBCO heterostructures is still under experimental debate \cite{fridman, chien}.

In this work, we distinguish between the effects of strain and magnetism on the superconductivity in $c$-axis LCMO/YBCO heterostructures, by studying the $T_c$ of perovskite/YBCO/perovskite trilayer thin films with either ferromagnetic LCMO or paramagnetic LaNiO$_3$ (LNO) as perovskite.  For comparison with samples that are lattice-symmetry matched, we also use orthorhombic PrBa$_2$Cu$_3$O$_{7-\delta}$ (PBCO) in place of the pseudocubic perovskites.  Both LCMO/YBCO/LCMO and LNO/YBCO/LNO trilayers show strong attenuation of $T_c$, as YBCO layer thickness is reduced from 21.4 to 5.4 nm, whereas PBCO/YBCO/PBCO trilayers show a much milder $T_c$ attenuation. These results indicate that heteroepitaxial strain, rather than long-range F/S proximity effect, is responsible for the long length scales of $T_c$ attenuation observed in $c$-axis LCMO/YBCO heterostructures.

\begin{table}[htb]
\centering
\begin{tabular}{|l|l|l|l|l|l|l|}
\hline
Material & Lattice & $a$ (\AA) & $b$ (\AA) & $\epsilon_{aa}$(\%) &$\epsilon_{bb}$(\%) \\ \hline
$\mathrm{YBa_2Cu_3O_{7}}$ & orthorhombic & 3.821 & 3.885 & $-$ & $-$ \\ \hline
$\mathrm{PrBa_2Cu_3O_7}$ & orthorhombic & 3.868 & 3.911 & +1.2 & +0.7 \\ \hline
$\mathrm{YBa_2Cu_3O_{6.35}}$ & tetragonal & 3.858 & 3.858 & +1 & $-$0.7 \\ \hline
$\mathrm{La_{2/3}Ca_{1/3}MnO_3}$ & pseudocubic & 3.858 & 3.858 & +1 & $-$0.7 \\ \hline
$\mathrm{LaNiO_3}$ & pseudocubic & 3.838 & 3.838 & +0.5 & $-$1.2 \\ \hline
\end{tabular}
\caption{Bulk lattice parameters for the oxides involved in this study. Oxygen-deficient YBa$_2$Cu$_3$O$_{6.35}$, which is tetragonal, is included for comparison. Columns 2 and 3 show $a$-axis and $b$-axis lattice parameters. Columns 4 and 5 show the $\%$ of lattice-parameter mismatch relative to $\mathrm{YBa_2Cu_3O_{7}}$.}
\end{table}

The trilayer thin films used in our study were grown on $c$-axis oriented $\mathrm{SrTiO_3}$ substrates by pulsed laser-ablated deposition (PLD). A KrF excimer laser was used, operating at 248 nm, 2 - 5 Hz and a fluence of 2 J/cm$^2$. The PLD growths were done at 750 - 800$^\circ$C in 200 mTorr of $\rm O_2$. After deposition, each films was annealed \textit{in situ} by slow cooling over 45 minutes to 300$^\circ$C in 760 Torr of $\rm O_2$ to optimally oxygenate the YBCO layer. For each trilayer sample, the YBCO layer was clamped symmetrically between either PBCO, LCMO or LNO layers.  For our resistance measurements, thickness of the clamping layers was fixed at 10.7 nm, while the YBCO layer was varied between 5.4 and 21.4 nm in $\sim$ 5.4 nm increments.  The range of relative thickness was deliberately chosen to tune the heteroepitaxial strain on the YBCO layer.

\begin{figure}[htb]
\centering
\includegraphics[width=3.3in]{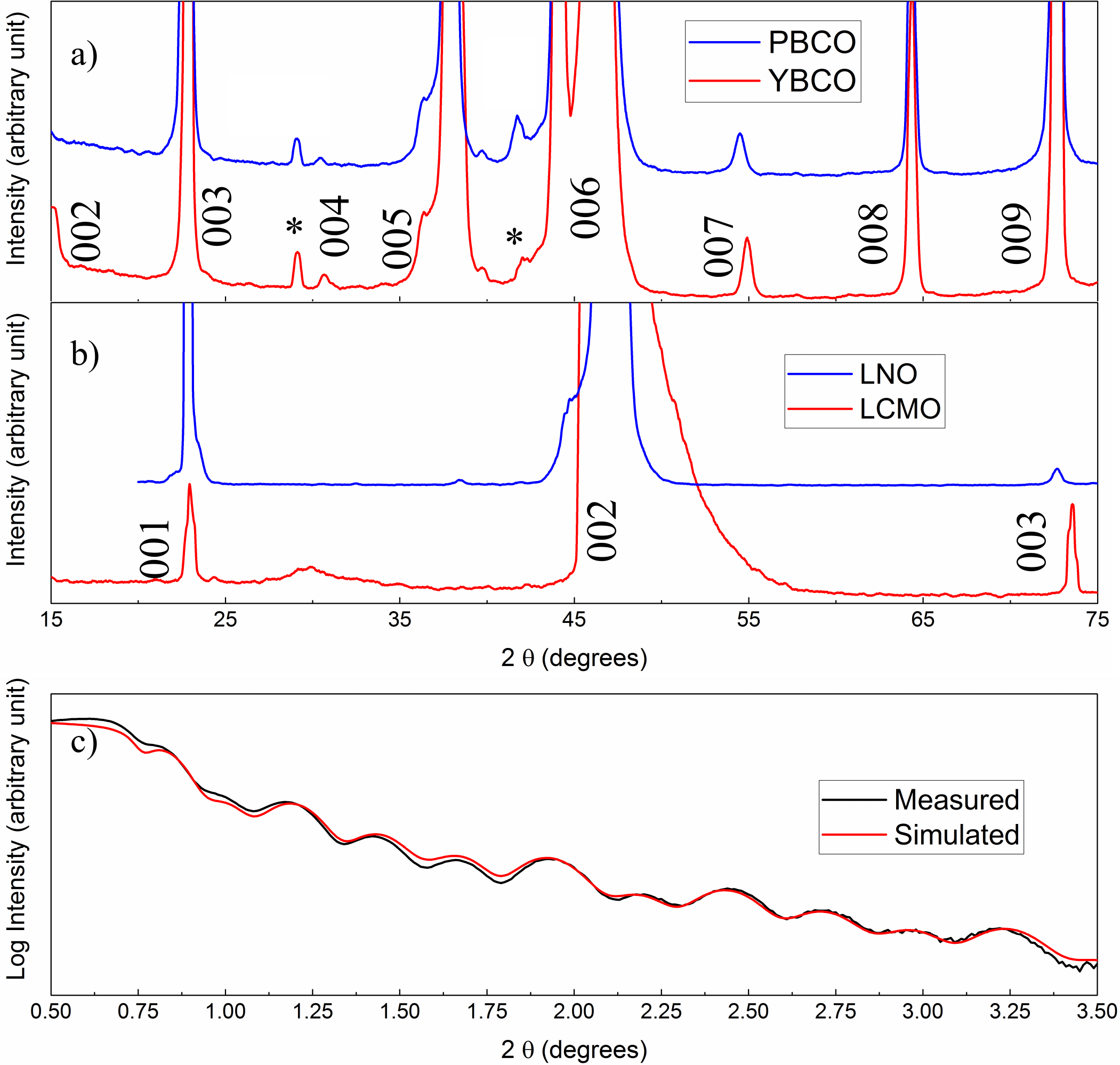}
\caption{Representative XRD and XRR patterns measured on our thin-film samples by $\theta-2\theta$ scans.  Panel (a) plots XRD data for unilayer YBCO and PBCO films, both being orthorhombic. Panel (b) plots XRD data for unilayer LCMO and LNO films, both being pseudocubic. The XRD patterns are similar for films with the same lattice structure. Panel (c) plots low-angle XRR data measured on a LNO/YBCO/LNO trilayer film, along with XRR data simulated by GenX refinement, showing good fit for model layer thicknesses of 12.7 nm/6.6 nm/12.7 nm and interfacial roughness of 4 \AA.
}
\end{figure}

Structural characterization of our thin-film samples was done by x-ray diffraction (XRD) and x-ray reflectometry (XRR), both using $\theta-2\theta$ scans.  The XRD data were taken with either a Bruker D8 DISCOVER or a Rigaku MiniFlex 600 diffractometer, and the XRR data were taken with a Rigaku Smartlab diffractometer. Fig. 1(a) and 1(b) plot the XRD data for unilayer films of YBCO, PBCO, LCMO and LNO, showing similar peak patterns for films with the same lattice structure.  The starred peaks are due to radiation contamination of W $L_\alpha$ and Cu $K_\beta$. Fig. 1(c) plots both the measured and simulated XRR patterns for a LNO/YBCO/LNO trilayer film made for growth rate calibration. The simulations were done with a simple trilayer model using the extensible XRR refinement program GenX \cite{fullerton, bjorck}.  For the XRR data shown in Fig. 1(c), the layer thicknesses were determined to be 12.7 nm/6.6 nm/12.7 nm, and the interfacial roughness between layers to be 4 \AA. Such XRR analysis provided accurate calibration for the layer thicknesses of our trilayer films. The growth rate for each material was also checked with atomic force microscopy, by measuring the thickness of a unilayer film after semi-masking and chemically etching it to create a step edge.  Finally, the electrical resistance of each film was measured versus temperature in a $^4$He cryostat, using standard ac lock-in technique in the four-contact configuration.

\begin{figure}[htb]
\centering
\includegraphics[width=3.3in]{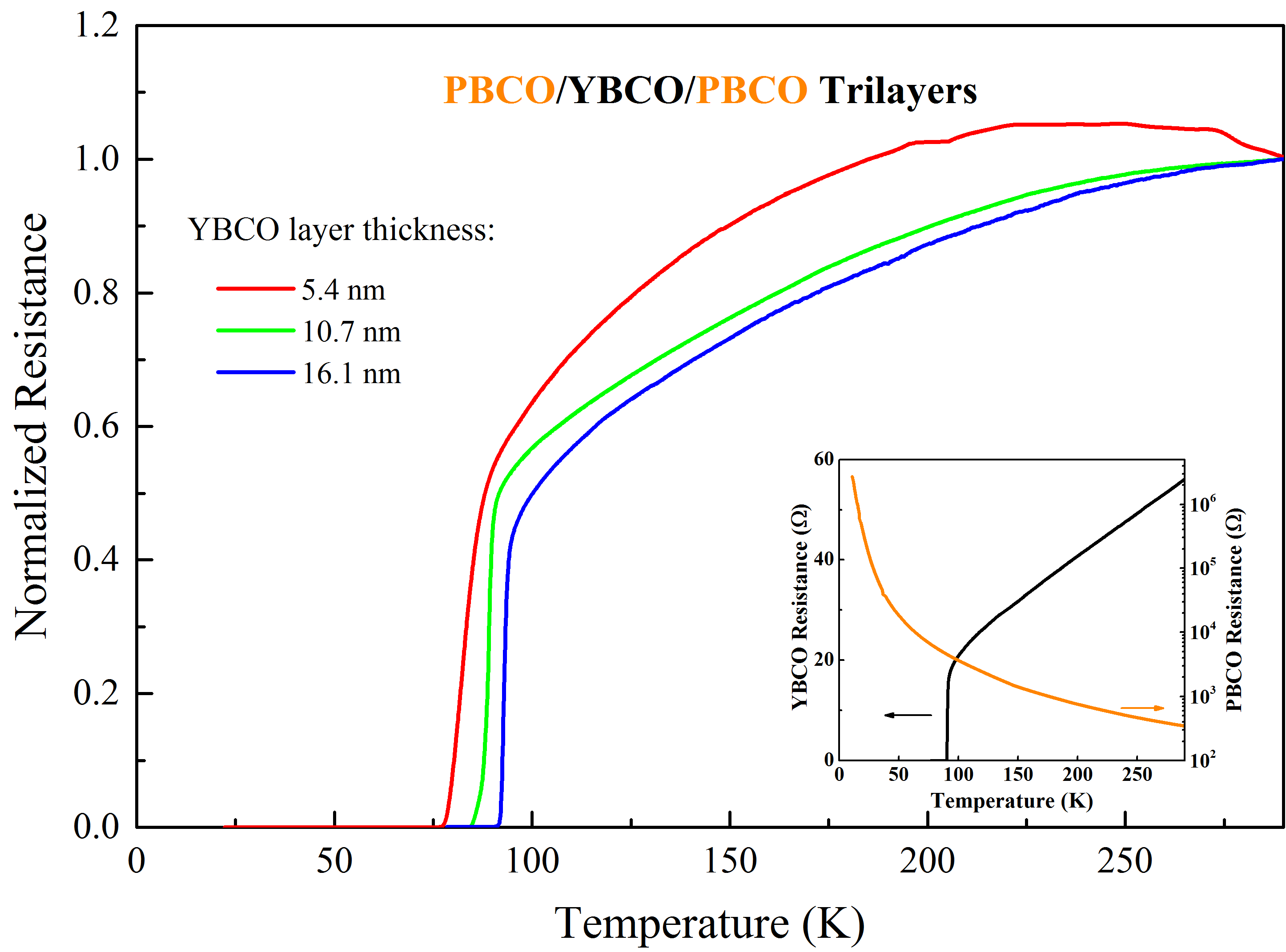}
\caption{Normalized resistance versus temperature of PBCO/YBCO/PBCO trilayer films, plotted for YBCO layer thickness ranging from 5.4 to 16.1 nm. All the films show sharp superconducting transitions with similar $T_c$ onsets near 90 K, attesting to the lattice-symmetry matching between YBCO and PBCO. The mild attenuation of $T_c$ vs. decreasing YBCO thickness can be understood in terms of the lattice-parameter mismatch, which stretches the $a$-axis of YBCO more than its $b$-axis, thus slightly reducing its oxygen content. For comparison, the inset shows data for unilayer YBCO and PBCO films that are 21.4 nm thick.}
\end{figure}

Figure 2 shows the resistance $R$ versus temperature $T$ data for PBCO/YBCO/PBCO trilayers of various YBCO thickness. To facilitate comparison, each $R$-vs.-$T$ curve is normalized by its room-temperature value. All the trilayers show sharp superconducting transitions, with the $T_c$ onset (defined as the intersection of two linear fits to the transition edge) being mildly attenuated from 94 to 91 and 88 K as the YBCO thickness is decreased from 16.1 to 10.7 and 5.4 nm. Similar $T_c$ attenuation was seen in previous studies of PBCO/YBCO/PBCO trilayers \cite{chan}, and can be explained by the PBCO/YBCO lattice-parameter mismatch, which stretches the $a$-axis of YBCO more than its $b$-axis (direction of the CuO chains), thus slightly reducing its oxygen content.  The mildness of this $T_c$ attenuation attests to the lattice-symmetry matching between PBCO and YBCO, in contrast to the LCMO/YBCO/LCMO and LNO/YBCO/LNO results presented below.  It is worth noting that at 5.4-nm YBCO thickness, there is a slope inflection near 210 K, which can be explained using a parallel-resistor model.  Namely, since PBCO is insulating, PBCO/YBCO/PBCO trilayers would appear insulating above/below $\sim$ 210 K when the YBCO layer is more/less resistive than the two PBCO layers in parallel.

\begin{figure}[htb]
\centering
\includegraphics[width=3.3in]{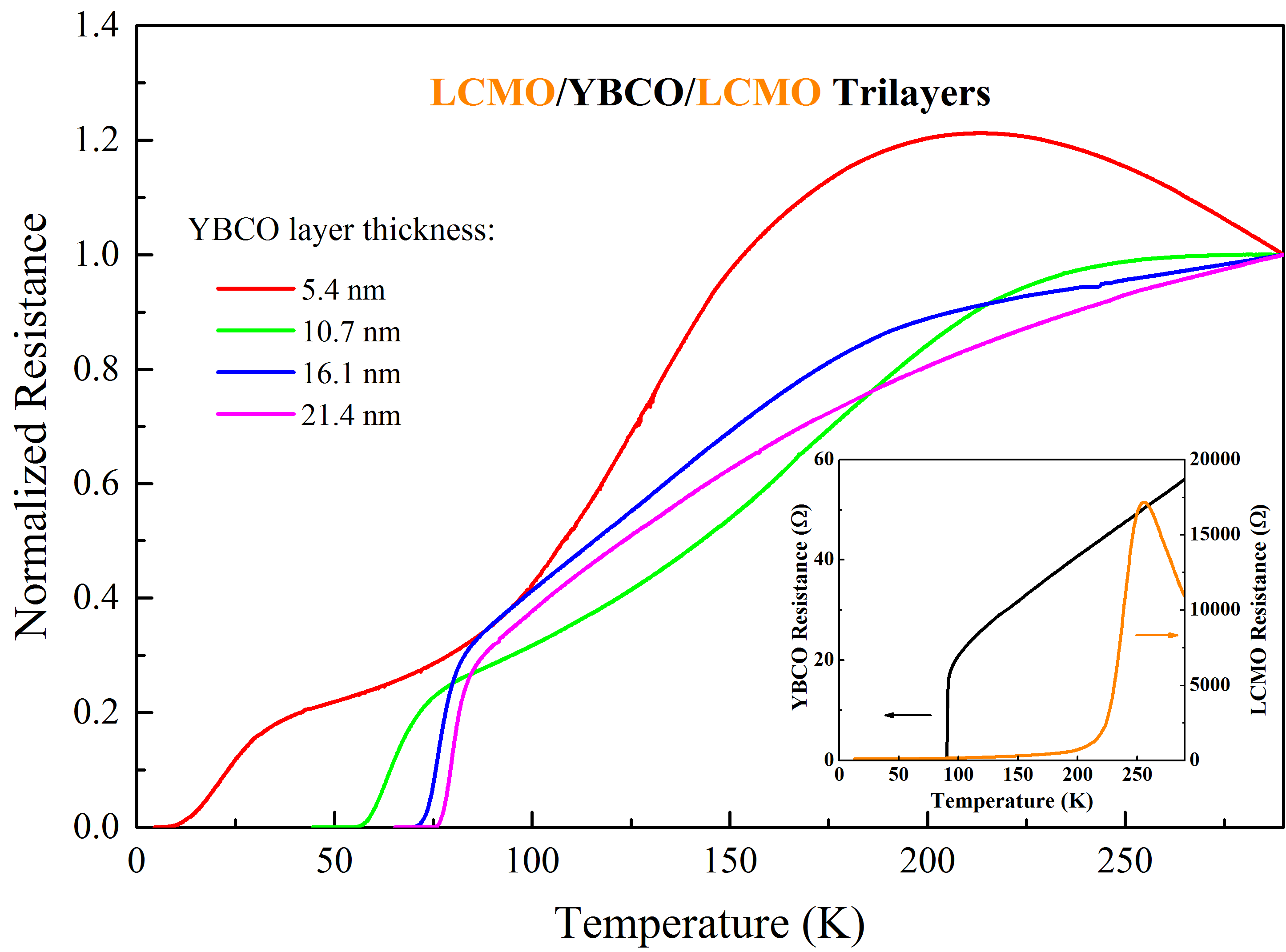}
\caption{Normalized resistance versus temperature of LCMO/YBCO/LCMO trilayer films. As YBCO thickness is decreased from 21.4 to 5.4 nm, the $T_c$ onset is strongly attenuated from 83 to 35 K, and the superconducting transition broadens.  This attenuation of superconductivity vs. decreasing YBCO thickness also occurs in LNO/YBCO/LNO (Fig. 4) but not in PBCO/YBCO/PBCO trilayers (Fig. 2). At 5.4-nm YBCO thickness, the slope inflection near 200 K can be explained using a parallel-resistor model for the trilayer, with the LCMO layer having a CMR peak near its Curie temperature. For comparison, the inset shows data for unilayer YBCO and LCMO films that are 21.4 nm thick.}
\end{figure}

Figure 3 shows the normalized $R$-vs.-$T$ data for LCMO/YBCO/LCMO trilayers of various YBCO thickness. As the YBCO thickness is decreased from 21.4 nm to 5.4 nm, the $T_c$ onset is strongly attenuated from 83 K to 35 K, and the superconducting transition broadens.  A similar $T_c$ attenuation was seen in our study of bilayer LCMO/YBCO films grown on $\mathrm{(La, Sr)(Al, Ta)O_3}$ (LSAT) substrates: the $T_c$ of a 25-nm YBCO film is significantly reduced when a 25-nm LCMO layer is grown over it \cite{zhang}. This $T_c$ reduction was attributed to the intergrowth of double-CuO chains, induced by heteroepitaxial strain from both the LCMO overlayer and the LSAT substrate. Since LCMO and LSAT are both pseudocubic in structure and similar in lattice parameter ($<$ 0.25 $\%$ mismatch), it is reasonable to assume that the strain effects on YBCO are similar whether it is clamped between LCMO or between LCMO and LSAT.  In our present study, it is worth noting that at 5.4-nm YBCO thickness there is a slope inflection near $\sim$ 200 K. This inflection can also be explained using a parallel-resistor model, with the LCMO layer having a colossal magnetoresistance (CMR) peak near its Curie transition.

\begin{figure}[htb]
\centering
\includegraphics[width=3.3in]{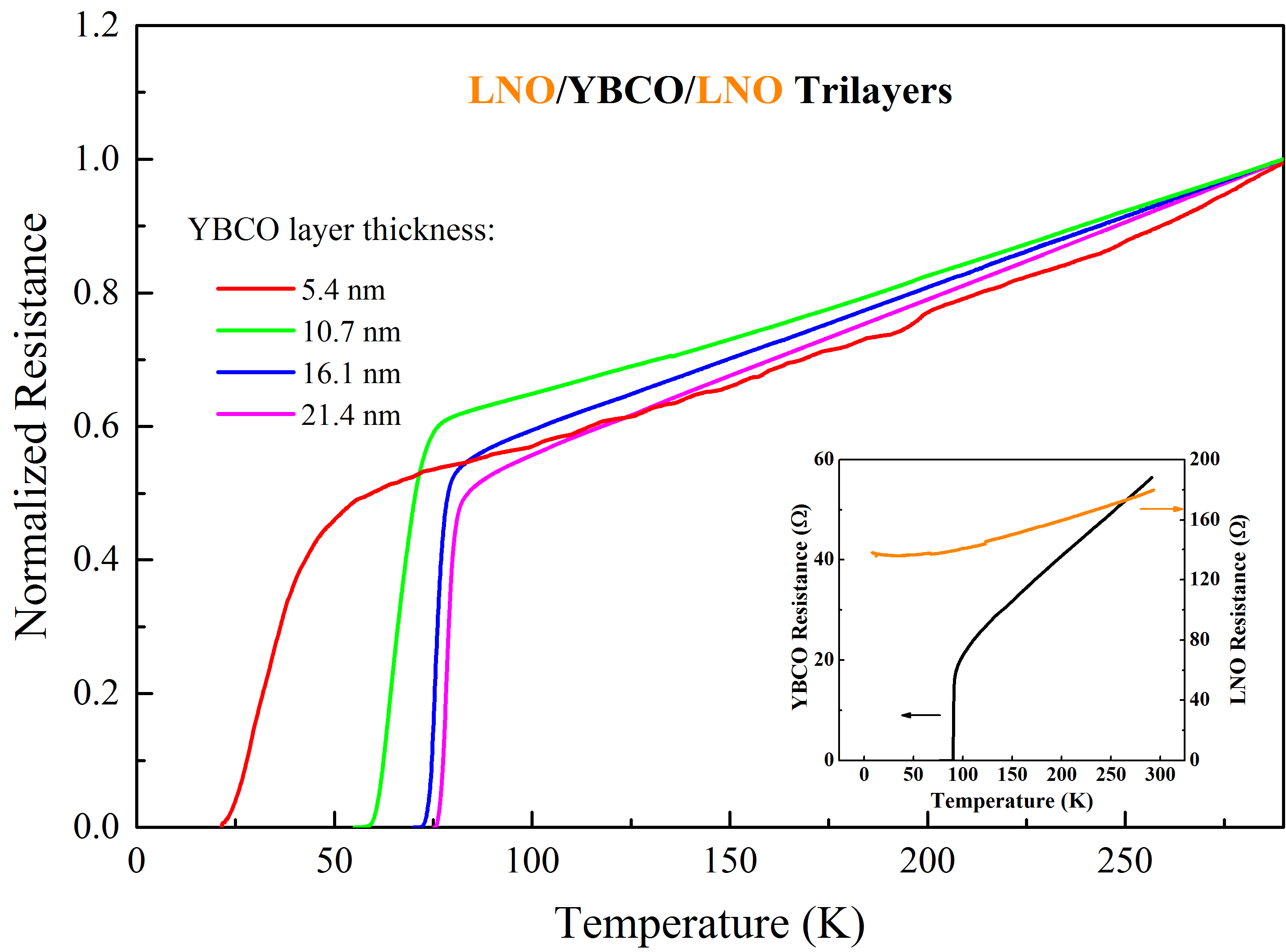}
\caption{Normalized resistance versus temperature of LNO/YBCO/LNO trilayer films. As YBCO thickness is decreased from 21.4 to 5.4 nm, the $T_c$ onset is strongly attenuated from 82 to 42 K, and the superconducting transition broadens. Apart from the absence of a CMR peak near 200 K and a sharper transition for the 5.4 nm case, the LNO/YBCO/LNO trilayers show similar behavior as their LCMO/YBCO/LCMO counterparts (Fig. 3). The similarities are illustrated in Fig. 5.  For comparison, the inset shows data for unilayer YBCO and LNO films that are 21.4 nm thick.}
\end{figure}

Figure 4 shows the normalized $R$-vs.-$T$ data for LNO/YBCO/LNO trilayers of various YBCO thickness. Apart from the absence of a CMR peak, all the LNO/YBCO/LNO samples show similar behavior as their LCMO/YBCO/LCMO counterparts (see Fig. 3). That is, the $T_c$ onset is strongly attenuated from 82 K to 42 K, and the superconducting transition broadens, as the YBCO thickness is decreased from 21.4 nm to 5.4 nm. In fact, except for the thinnest (5.4-nm) YBCO case, the $T_c$ onsets and transition widths agree to within 2 K between the LNO/YBCO/LNO and LCMO/YBCO/LCMO data. This agreement is illustrated in Figure 5, which plots $T_c$ versus YBCO layer thickness for all three types of trilayers measured.  The similarity between the LNO/YBCO/LNO and LCMO/YBCO/LCMO data indicates that the $T_c$ attenuation is less associated with magnetism of the perovskites than with heteroepitaxial strain between these pseudocubic perovskites and orthorhombic YBCO.  The contrast to the PBCO/YBCO/PBCO data is consistent with the lattice-symmetry matching between YBCO and PBCO, as described earlier.  At 5.4-nm YBCO thickness, it is possible that the lower $T_c$ onset and broader transition for LCMO/YBCO/LCMO vs. LNO/YBCO/LNO are due to F/S proximity coupling between LCMO and YBCO.  However, this length scale is much shorter than the purported range ($\sim$ 10 - 20 nm) of the F/S proximity effect for $c$-axis LCMO/YBCO heterostructures \cite{sefrioui, soltan, pang, pena}, thus still calling this long range into question.

\begin{figure}[htb]
\centering
\includegraphics[width=2.85in]{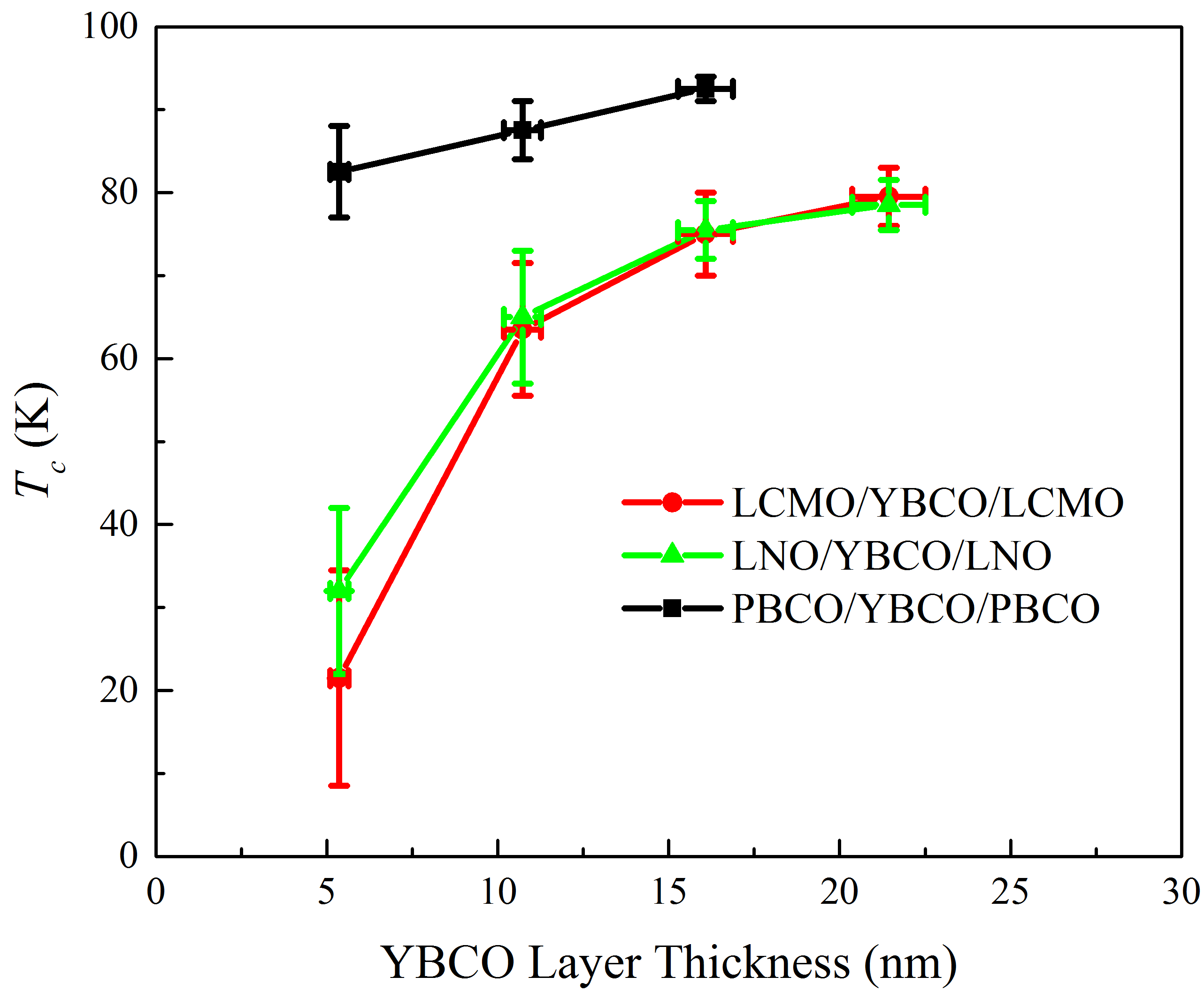}
\caption{Plot of $T_c$ versus YBCO layer thickness for all three types of trilayer films measured. Top/bottom end of each error bar corresponds to onset/completion of the superconducting transition. The similarity between LNO/YBCO/LNO and LCMO/YBCO/LCMO indicates that the $T_c$ attenuation is less associated with magnetism of the perovskite than with heteroepitaxial strain between these perovskites and the orthorhombic YBCO. The contrast to the PBCO/YBCO/PBCO data is consistent with the lattice-symmetry matching between YBCO and PBCO.}
\end{figure}

A closer examination of the lattice parameters listed in Table 1 sheds further light on the various dependences of $T_c$ versus YBCO thickness shown in Fig. 5.  Whereas the heteroepitaxial strain of PBCO on YBCO is consistently tensile in the $ab$-plane, the strain of either LCMO or LNO on YBCO is anisotropic, being tensile ($\epsilon_{aa}>0$) along its $a$-axis but compressive ($\epsilon_{bb}<0$) along its $b$-axis.  As shown in our previous study \cite{zhang}, this anisotropic strain can induce nanoscale formation of the $\mathrm{Y_2Ba_4Cu_7O_{15-\delta}}$ (YBCO-247) phase, which is less orthorhombic and tends to have lower $T_c$ than $\mathrm{YBa_2Cu_3O_{7-\delta}}$ \cite{morris,kato,irizawa}. Even if the YBCO-247 phase does not robustly form, the remaining $\mathrm{YBa_2Cu_3O_{7-\delta}}$ can still relieve the anisotropic strain by losing oxygen from its CuO chains, which run along the compressed $b$-axis, thereby lowering $T_c$. Such a strain-relieving mechanism via formation of oxygen vacancies has been observed in a variety of oxides \cite{klie, aschauer}. For oxygen-deficient $\mathrm{YBa_2Cu_3O_{6.35}}$, which is tetragonal, its in-plane lattice parameter (3.858 \AA) is in fact the same as that of LCMO. Thus it is inherently unfavorable for the YBCO within LCMO/YBCO/LCMO trilayers to remain strongly orthorhombic and fully oxygenated.  In essence, by heteroepitaxially clamping orthorhombic YBCO between pseudocubic perovskites in thin-film form, one can no longer ensure that the YBCO layer retains either its phase purity or oxygen stoichiometry.

The above analysis implicates the anisotropic strain due to lattice-symmetry mismatch as driving the attenuation of superconductivity in $c$-axis LCMO/YBCO heterostructures.  Although strain was reported to affect the ferromagnetic modulations seen in Pr-doped LCMO/YBCO superlattices \cite{hoppler}, the lattice-symmetry mismatch between LCMO and YBCO is a crucial experimental issue that has generally been overlooked. In addition to inducing ordered defects in the YBCO lattice which can affect $T_c$ \cite{zhang}, this symmetry mismatch may also affect the novel types of charge-density-wave order recently observed in LCMO/YBCO bilayers and superlattices \cite{he,frano}, by affecting either the electronic nematicity or the interplay between multiple order parameters \cite{comin,gerber,hayward}. Further studies correlating structural, transport and spectroscopic measurements are called for to elucidate these possibilities. For our present study of trilayers, the predominant effect of strain over magnetism implies that the long length scales of $T_c$ attenuation observed in $c$-axis LCMO/YBCO heterostructures cannot be identified with the F/S proximity effect. To avoid the anisotropic strain, the tetragonal La$_{2-x}$Sr$_x$CuO$_4$ (LSCO) could be used instead of orthorhombic YBCO \cite{koren,das,deluca}. Specifically, a comparative study of $c$-axis LCMO/LSCO/LCMO and LNO/LSCO/LNO trilayers, both being interfacially-matched in lattice symmetry, would enable a clearer determination of the range of the $c$-axis F/S proximity effect in hole-doped manganite/cuprate heterostructures.

In summary, we have studied $c$-axis perovskite/YBCO/perovskite trilayer thin films, using either ferromagnetic LCMO or paramagnetic LNO as the perovskite, to distinguish between the effects of strain and magnetism on the superconductivity.  LCMO/YBCO/LCMO and LNO/YBCO/LNO trilayers show similarly strong attenuation of $T_c$, as YBCO layer thickness is reduced in the range of 21.4 to 5.4 nm.  PBCO/YBCO/PBCO trilayers, which are lattice-symmetry matched, show a much milder $T_c$ attenuation. Our results indicate that heteroepitaxial strain, rather than long-range F/S proximity effect, is responsible for the long length scales of $T_c$ attenuation observed in $c$-axis LCMO/YBCO heterostructures.

This work was supported by NSERC, CFI-OIT and the Canadian Institute for Advanced Research through the Quantum Materials Program.

\end{document}